\documentstyle[preprint,revtex,eqsecnum]{aps}
\begin{document}
\draft
\preprint{UAHEP-935}
\begin{title}
Conjectures on Non-Local Effects in String Black Holes
\end{title}
\author{B.Harms and Y.Leblanc}
\begin{instit}
Department of Physics and Astronomy, The University of Alabama\\
Box 870324, Tuscaloosa, AL 35487-0324
\end{instit}
\begin{abstract}
We consider the modifications to general relativity by the non-
local (classical and quantum) string effects for the case of
a D-dimensional Schwarzschild black hole.  The classical
non-local effects do not alter the spacetime topology
(the horizon remains unshifted, at least perturbatively).
We suggest a simple analytic
continuation of the perturbative result into the non-
perturbative domain, which eliminates the black hole singularity
at the origin and yields an ultraviolet-finite
theory of quantum gravity.  We investigate the quantum non-local effects
(including massive modes) and argue that the inclusion of these back
reactions resolves the problem of the thermal spectrum in the semi-
classical approach of field quantization in a black hole background,
through the bootstrap condition. The density of states for both the
quantum and thermal interpretation of the WKB formula are finally shown
to differ quantitatively when including the non-local effects.
\end{abstract}

\narrowtext
\section{Introduction}

In previous works ~\cite {BL92,BL93,HL93a,HL93b}, we
have shown that the semiclassical (WKB) approximation
of the Euclidean path integral of a quantized theory
of gravity in a black hole background should properly be interpreted
as the probability of quantum tunneling of a particle across the black
hole horizon barrier. This interpretation yielded directly the quantum
black hole degeneracy of states and resolved all the problems associated
with the traditional thermodynamical interpretation in a way consistent
with the laws of thermodynamics and quantum mechanics. We were then led
to the "finger-printing" of D-dimensional neutral black holes through
their degeneracy of states and to their identification as quantum
$(D-2)\over(D-4)$-branes.

Although our work has shed some light on the quantum black hole problem,
still questions remain unanswered. Of particular urgency is the classic
problem of field quantization in the classical black hole background.
It has been demonstrated over and over again that ``naive
quantization'' of fields in such
a background leads to an {\it exact} thermal
distribution for the particle number density with the (Hawking)
temperature related to the mass of the black hole, in agreement with
the WKB approximation calculation.

It has often been argued, and extensively investigated in various
models and dimensions, that quantum gravity effects (loop corrections
or back reaction effects) would actually resolve the problem of the
thermal spectrum. That it is indeed a problem is related to the
prediction that a black hole having absorbed initially particles in
pure states will emit them at later times in a thermal (mixed) state
during its evaporation process, thereby violating the law of quantum
mechanics asserting that pure states evolve solely into pure states
in any time-dependent process.

In this paper, we intend to investigate the nature of the above problem
through a series of (qualitative) arguments, partially supported
by explicit computations.

It is important to get a headstart on this problem and its resolution
by first accepting a present-day fact about quantum gravity, namely,
the only known "consistent" theories of quantized gravity are those
of quantum p-brane theories. In this work, as a simple working model,
we shall investigate critical bosonic string (1-brane) theory
compactified down to D dimensions. With the acceptance of such theories
as quantum gravity theories, comes the realization that general
relativity (GR) now becomes doubly modified, once through higher
derivative (non-local) classical contributions arising fundamentally
from the intrinsic non-local nature of p-branes, and once from purely
quantum (loop) corrections, the so-called back-reaction effects,
which must include the higher massive string (or p-brane) excitations.
In this work we study the latter qualitatively and the former
quantitatively.

Non-local effects, such as (Riemann) curvature-square corrections
to Einstein's equations, have been considered by
previous authors ~\cite{green,Cal88,Myers}.
We review such calculations in the following
section by applying perturbation theory about the D-dimensional
Schwarzschild black hole metric. The small perturbation parameter
is taken to be the Regge slope $\alpha'\,=\,{1\over{2\pi{T}}}$,
where $T$ is the string tension. The non-local $\alpha'$-corrections
were found ~\cite{Cal88} not to modify Hawking's
black hole thermodynamics in any significant way. The horizon
(topological) structure of the black hole spacetime remains undisturbed
by these non-local effects, at least perturbatively.

Pending plausible analytical continuations to strong $\alpha'$
($\alpha'\,\rightarrow\,\infty$), we show that the black
hole singularity at the origin can be eliminated. So,
non-local effects remove the ultraviolet
divergences of quantum gravity in all Feynman loop diagrams involving
graviton propagators, a purely stringy effect directly
related to the removal of the black hole singularity
($r\,\rightarrow\,0$). With such corrections, gravity is shown to
vanish at the black hole center (asymptotic freedom). Although
the Hawking temperature is in general
modified by the $\alpha'$-corrections,
Green's functions of quantized fields in the modified black hole
background remain periodic with respect to the (modified) inverse
Hawking temperature (Kubo-Martin-Schwinger property) and so again
lead to an exact thermal spectrum for the particle number density,
even though the black hole singularity is removed.

Perturbatively, the classical stringy corrections to Einstein's
equations do not disturb the global topology of the spacetime.
This illustrates the point that the seemingly thermal property of
black holes is not intrinsic to the black hole as a curvature-
singular object. In fact, any regular spacetime, whether curved or
not and which possesses an horizon, leads to an exact thermal
spectrum (e.g. static de Sitter space, Rindler space, etc.).

Clearly the resolution of the problem of the exact thermal spectrum
{\it must} lie within the quantum effects. Fundamentally, the origin
of the Hawking's radiation (thermal spectrum) phenomenon
lies in the fact that the horizon acts
as an impenetrable barrier between classically
causally disconnected sectors of spacetime and that, although a quantized
field may exist everywhere in the spacetime, information on the field
from sectors not attainable by an observer is seemingly lost.

Finding the true vacuum is a highly non-trivial exercise and is truly
a self-consistent (nonperturbative) problem in which the classical
background determines the quantization process which itself determines
this same background.

In the last section, we show how the bootstrap condition inherent in
the quantum (tunneling probability) interpretation of the WKB formula
leads to a possible (non-perturbative) solution of the thermal spectrum
problem.

\narrowtext
\section{String Corrections to Black Hole Backgrounds}

Corrections to the metric tensor due to `stringy' (non-
local) classical effects were carried out to $O(\alpha')$ in
Ref.\cite{Cal88}.  For a bosonic string the usual form for
the action
\begin{eqnarray}
S_0 = {1\over{16\pi}}\int d^Dx\sqrt{-g}e^{-2\phi}\bigl(R +
4(\nabla\phi)^2 + {\alpha'\over{4}}R_{klmn}R^{klmn}\bigr)\;
,
\end{eqnarray}
where $D$ is the dimension of the black hole, gives the
equations of motion for the graviton and the
dilaton
\begin{eqnarray}
R_{ij} + 2\nabla_i\nabla_j\phi +
{\alpha'\over{2}}R_{iklm}R_j^{klm} = 0\; ,\\
\raisebox{.6ex}{\fbox{\rule{0mm}{0mm}}}\;\phi -
(\nabla\phi)^2 +
{1\over{4}}R
+
{\alpha'\over{16}}R_{klmn}R^{klmn} = 0 \; .
\end{eqnarray}
The action given in Eq.(2.1) leads to a difference between the
inertial and gravitational masses of the black hole due to the
coupling of the dilaton to the trace of the graviton.  This
problem can be overcome by making the conformal
transformation ~\cite{Cal88},
\begin{eqnarray}
g_{ij} \to \exp({4\over{D-2}}\phi)\;g_{ij} \; .
\end{eqnarray}
This transformation, together with some field
redefinitions, changes the action to,
\begin{eqnarray}
S' = {1\over{16\pi}}\int d^Dx\sqrt{-g}\Bigl( R - {4\over{D-
2}}(\nabla\phi)^2 + {\alpha'\over{4}}e^{-4\phi/(D-
2)}R_{klmn}R^{klmn}\Bigr)\; .
\end{eqnarray}
The resulting equations of motion
\begin{eqnarray}
R_{ij} + {\alpha'\over{2}}e^{-4\phi/(D-2)}\Bigl(
R_{iklm}R_j^{klm}
- {1\over{2(D-2)}}\;g_{ij}\; R_{klmn}R^{klmn}\Bigr) = 0\; ,
\nonumber
\\
\raisebox{.6ex}{\fbox{\rule{0mm}{0mm}}}\;\phi -
{\alpha'\over{8}}e^{-
4\phi/(D-2)}
R_{klmn}R^{klmn} =
0\; ,
\end{eqnarray}
can be solved perturbatively by expanding about the $\alpha'
= 0$ Schwarzschild metric for a static black hole and about
the constant part of the dilaton field.  To first order in
$\alpha'$ the elements of the metric tensor and the dilaton
field are given by
\begin{eqnarray}
g_{00} = g^{(0)}_{00}(1 + {\alpha'}f(r)) \; ,
\nonumber \\
g_{11} = g^{(0)}_{11}(1 + {\alpha'}g(r)) \; , \nonumber \\
g_{ii}{d}{x_i^2} = r^2d\Omega^2_{D-2}\;;\ \  i \geq 2, \\
\phi = \phi_0 + {\alpha'\over{2}} \varphi (r)\; ,\nonumber
\end{eqnarray}
where
\begin{eqnarray}
g^{(0)}_{00} = (-g^{(0)}_{11})^{-1} = -(1-{\kappa\over{r^{D-
3}}})\; ,
\end{eqnarray}
$d\Omega^2_{D-2}$ is a line element on the unit $D-
2$-sphere, and $\phi_0$ is an arbitrary constant.  $\kappa$
is related to the unperturbed mass of the black hole
by,
\begin{eqnarray}
m = {(D-2)A_{D-2}\over{16\pi}}\kappa\; ,
\end{eqnarray}
where $A_D$ is the area of a unit D-sphere.
Using these
expressions in Eq.(2.6) it is found that ~\cite{Cal88},
\[
f = -g \; ,
\]
\begin{eqnarray}
g = {(D-3)(D-4)\over{4}}\;{r_+^{D-5}\over{r^{D-1}}}\;{r^{D-
1}-
r_+^{D-1}\over{r^{D-3}-r_+^{D-3}}} \; ,
\end{eqnarray}
and $\varphi$ satisfies the differential equation
\begin{eqnarray}
{d\varphi\over{dr}} = {(D-2)^2(D-3)\over{4}}\;{r_+^{D-
5}\over{r^D}}\;{r^{D-1} - r_+^{D-1}\over{r^{D-3}-r_+^{D-
3}}}\; ,
\end{eqnarray}
$r_+$ being the value of $r$ at the horizon, $r_+ =
\kappa^{{1\over{D-3}}}$.
These expressions show that to first order in $\alpha'$
there are no corrections to the metric tensor elements in four
dimensions.  However in higher dimensions and for higher
orders
in the $\alpha'$ expansion in four dimensions the metric
tensor is modified and
hence the action and the thermodynamical quantities
calculated from it are also modified.

\narrowtext
\section{String Corrections to the Thermodynamics of Black
Holes}

Non-local contributions to the metric tensor will affect the
thermally interepreted character of the geometry.  According
to Gibbons and Hawking \cite{gibb}, the thermodynamical
properties of a black hole with inverse temperature
$\beta_H$ can be obtained by analytically continuing the
metric to Euclidean spacetime, and summing over geometries
which are asymptotically flat, have topology $R^2\times S^2$
and are periodic in imaginary time $\tau$ with period
$\beta_H$.  In bosonic string theory the spacetime dimension
is 26, so that we consider a manifold which is a direct
product of a $D$-dimensional black hole and a $26-D$-
dimensional compact internal space.  The thermodynamical
quantities of interest are obtained from the partition
function, which in the usual semiclassical (WKB)
approximation is given by,
\begin{eqnarray}
Z(\beta_H) \sim \exp(-S_E/\hbar)\; ,
\end{eqnarray}
where $S_E$ is the Euclidean action evaluated at the black
hole solution.  The total Euclidean action is the sum of the
analytically continued action of Eq.(2.5) and a boundary
term $S_{bd}$ (see Ref.[6]).

As usual $\beta_H$ is the inverse Gibbons-Hawking temperature
given by,
\begin{eqnarray}
\beta_H = {2\pi\over{[(\partial_r e^{\Phi(r)})e^{-
\Lambda(r)}]_{r=r_+}}} \; .
\end{eqnarray}
where the
functions $\Lambda(r)$ and $\Phi(r)$ are given by
\begin{eqnarray}
\Lambda(r) = {1\over{2}}\ln (g_{11}) \; ,\nonumber \\
\Phi(r) = {1\over{2}}\ln (g_{00}) \; .
\end{eqnarray}
$S_{bd}$ is to be evaluated at spatial infinity but actually
diverges as $r\to\infty$. The corresponding action
of a flat spacetime at the boundary must be subtracted
\cite{gibb} to eliminate this divergence.  The total
Euclidean action is
\begin{eqnarray}
S_E &=& S+S_{bd}-S_{bd}^{flat} \;, \nonumber \\
&=& {A_{D-2}\over{16\pi}}\; \beta_H r_+\Bigl( 1-(D-2)(D-
3){\alpha'\over{2r_+^2}}\Bigr) \; .
\end{eqnarray}

To order $\alpha'$ the inverse temperature $\beta_H$ is
given by
\begin{eqnarray}
\beta_H = {4\pi r_+\over{D-3}}\;\Bigl( 1+{(D-2)(D-
4)\over{2}}\; {\alpha'\over{2r_+^2}}\Bigr)\; .
\end{eqnarray}
The entropy to the same order in $\alpha'$ is
\begin{eqnarray}
S &=& \beta_H\langle E\rangle - S_E \nonumber \\
&=& {A_{D-2}\over{4}}r_+^{D-2}\Bigl( 1+{(D-
2)^2\over{2}}\;{\alpha'\over{2r_+^2}}\Bigr) \; .
\end{eqnarray}
where $\langle E\rangle = M$, the $\alpha'$-corrected mass
of the black hole ~\cite{Cal88},
\begin{eqnarray}
M\;=\;m[1\,+\,{{{\alpha'}(D-3)(D-4)}\over{4{r_+^2}}}]\;.
\end{eqnarray}
Eq.(3.6) shows that stringy effects modify the
area law. However the first
law of black hole thermodynamics
\begin{eqnarray}
\beta_H = {dS\over{dM}} = 8\pi M \; ,
\end{eqnarray}
is still satisfied.

\narrowtext
\section{Ultraviolet Finite Quantum Gravity}

The content of this section is primarily based upon the Limiting
Curvature Hypothesis (LCH) ~\cite{Brand}
which states that no classical curvature
singularity should occur in a suitable geometrical gravitational
theory. This is required in order for the spacetime to be geodesically
complete. This principle effectively provides for the realization
of Penrose's Cosmic Censorship Hypothesis.

Another important input into the working assumptions of this section
is the belief that the non-local p-brane theories (including of
course string theories) are ultraviolet (UV) finite theories.

As is clear from the viewpoint of the solution Eqs.(2.7),(2.10)
for the black hole metric incorporating the non-local effects to first
order in $\alpha'$, the singular behavior at the origin of the
metric worsens as we go to higher orders in $\alpha'$-perturbation
expansion. That it is so originates from the fact that the higher order
terms in the corrected Einstein's equations have an increasingly
higher number of derivatives. The perturbation series in $\alpha'$
is therefore a series of individually divergent terms as $r \rightarrow
0$. However, in view of the arguments of the preceding paragraphs,
the series itself should be regular as $r \rightarrow 0$. To uncover
the full solution would probably require a great deal more calculation
involving the higher order terms.

For our purposes however, a simple analytic continuation of the
perturbative results to strong $\alpha'$ (although it is by no means
unique) will suffice. Let us consider the following analytical
continuation to strong $\alpha'$ (or small $r$),
\begin{eqnarray}
g_{oo}\;=\;-1\;+\;{{r_+^{D-3}}\over{r^{D-3}}}\,
{\exp}[{\alpha'}C({1\over{r_+^{D-1}}} - {1\over{r^{D-1}}})]\;.
\end{eqnarray}
where we defined $C\,\equiv\,{{(D-3)(D-4){r_+^{D-3}}}\over{4}}$.

Notice that the horizon of this spacetime is again left undisturbed
by the $\alpha'$-effects.

It is now apparent that the singularity at the origin has been removed.
Indeed spacetime is flat at the origin.
Now since $g_{oo}\,=\,-1 - 2U(r)$,
in which $U(r)$ is the gravitational potential, Eq.(4.1) gives the
following stringy modification to Newton's gravity,
\begin{eqnarray}
U(r)\;\propto\;- {{M}\over{r^{D-3}}} \exp[{{{-\alpha'}
C}\over{r^{D-1}}}]\;.
\end{eqnarray}
It may be useful to comment here that the non-local $\alpha'$-corrections
to Newton's law will produce modifications (although very small) to
phenomena such as the gravitational redshift, the bending of light by
the sun and the perihelion of Mercury. If refined experiments were
carried out (which may be decades or centuries away), these would allow
for the first direct measurements of the string tension.

Let us now express the Fourier transform of
the gravitational potential $U(r)$ as follows,
\begin{eqnarray}
U(k)\;\equiv\;- {M\over{A_{D-2}}}
{\int_{-\infty}^{\infty}} d^{D-1}{\vec x}
\bigl\{ {{|\vec x|}^{3-D}} \exp{\bigl[{{-{\alpha'}
C}\over{|\vec x|^{D-1}}}
\bigr]} \bigr\} e^{i{\vec k}\cdot{\vec x}}\;\;;
(k = |\vec k|)\;.
\end{eqnarray}
We find,
\begin{eqnarray}
U(k)\;=\;- M \lim_{\epsilon \rightarrow {0}} \int_0^{\infty} dr
r^{1-\epsilon} {\sin{kr}\over{kr}} \exp{\bigl[{{-{\alpha'}C}
\over {r^{D-1}}}\bigr]}\;,
\end{eqnarray}
in which $\epsilon$ is an infrared (IR) ($r\rightarrow\infty$) regulator.

Introducing the dimensionless parameter $\sigma \equiv kr$, we finally
arrive at the following explicit integral representation for the
graviton ``propagator'' $U(k)$,
\begin{eqnarray}
U(k)\;=\;- {M\over{k^2}} \lim_{\epsilon
\rightarrow 0} \int_0^{\infty} d\sigma\,{{\sin{\sigma}}\over{\sigma^
{\epsilon}}}\,exp[{{-{\alpha'}Ck^{D-1}}\over{\sigma^{D-1}}}]\;.
\end{eqnarray}

Two important remarks must be made. First, as is clear from the
above formula, the IR behavior of the graviton propagation function
remains unaltered by the (re-summed) non-local $\alpha'$-effects.
The graviton remains a massless particle and no ghost particles occur
in this expression. The second remark of course concerns the UV
behavior of the propagator, which clearly vanishes exponentially
as $k \rightarrow \infty$, at least for $D>4$. In 4 dimensions,
the corrections to the propagator are of
order {\it O}(${\alpha'}^2$) ~\cite{Myers}.
It is now obvious that any loop Feynman
diagrams involving internal graviton propagators lead to UV-finite
results. Loop diagrams should be evaluated by first integrating
over the loop momenta, then integrating over the integral representation
parameters $\sigma_i$'s and finally taking the $\epsilon_i$'s IR
regulators to zero, in that order.

Of course these apply to the Feynman rules of a field theory of gravity,
namely quantum GR, which itself is the ${\alpha'}\rightarrow 0$ limit
of string theory. The above considerations show that not all
${\alpha'}$-effects should be taken to zero in order to recover the
field theory limit, if finite predictions are to be made from such
a theory.

\narrowtext
\section{Stringy Back Reaction Effects}

Up to this point, we addressed the problem of the effects of the
classical non-local ($\alpha'$-corrections) modifications of GR from
the string progenitor, in connection with black hole
spacetimes.

The results are that the non-local effects do not disturb the topological
character of the spacetime  (the horizon remains unshifted, at least
perturbatively).

In addition, pending the plausible analytical continuations of the
results of Section II to the strong $\alpha'$ domain, as discussed in
Section IV, we found that the quantum field theory limit (quantum GR)
of string theory is UV-finite, provided some $\alpha'$-dependency
remains in the field theory limit.

On the other hand, because the non-local classical $\alpha'$-corrections
do not influence the topology of the black hole (although they do
generally reduce its Hawking
temperature), field quantization in the black
hole background will continue to yield a thermal spectrum (with
shifted Hawking temperature) for the particle number density, a
somewhat disappointing result if one is to believe in the pure quantum
nature of black holes.

As mentioned in the Introduction, the resolution of the
thermal spectrum must originate from the quantum or back reaction
effects. Attempts to resolve this problem have been made by various
authors and in various numbers of dimensions.

An intuitive, if not totally naive, way to see that this may indeed
occur, is the following consideration making use of the statistical
bootstrap model of quantum Schwarzschild
black holes ~\cite{BL92,BL93}. The situation
is depicted in Fig.~\ref{boot}, where a single massive quantum black hole
is modeled as a gas consisting itself of a single quantum massive
black hole excitation surrounded by countless extreme (massless)
others. Of course the equilibrium state of this gas is not thermal,
as the energy distribution is highly
inhomogeneous ~\cite{BL92,BL93}. If for some
reason the massive excitation is cut-off from the rest, the original
quantum black hole will resemble a gas of massless excitations
in thermal equilibrium. It is the effect of the single massive
excitation in the gas which drastically changes the nature of the
equilibrium configuration and allows
for the bootstrap property. This massive
excitation of course is a pure quantum gravity (back reaction)
effect of the p-brane type. Should such effects be included in the
usual semiclassical treatments, the above statistical model
shows that the thermal spectrum problem would be resolved
satisfactorily by yielding a picture of the original black hole
as a pure state. It must be pointed out that the above possible
resolution of the thermal spectrum problem is fundamentally non-
perturbative as one cannot separate linearly classical and quantum
effects in the WKB semiclassical approximation. This seems to imply
that the usual field quantization in the black hole background
may become fully consistent only after resummation of a certain
class of Feynman diagrams. This is tantamount to trying to reach
the superconducting phase by perturbing about the normal phase.
Obviously, this type of calculation is going nowhere.

As a final comment, it may very well be sensible
to expect quantum gravity
effects to induce an effective cosmological constant in the original
metric as loop corrections induce a non-zero vacuum energy.
Consequently, it is possible that there is no such a thing as
a quantum Schwarzschild black hole. The metric instead would be of the
Schwarzschild-de Sitter type, which is known to contain at most two
real horizons. This quantum-induced change of the topology of the
black hole spacetime may have drastic effects on the thermal
distribution of the particle number density already at the level
of loop perturbation theory. Explicit computations
are currently being carried out.

\narrowtext
\section{Conclusion}

In this work, we have looked at the modifications to the D-dimensional
Schwarzschild black hole solution of GR due to stringy $\alpha'$-effects,
either classically or quantum mechanically. We have presented conjectures
on how non-perturbative $\alpha'$-effects might affect classically
the black hole singularity at the origin and thus lead to its removal, as
well as how non-perturbative (WKB) quantum effects might resolve the
problem of the thermal spectrum. In the former case, the argumentation
was quantitative in nature while the latter case was
treated qualitatively.

As a final comment, let us point out that, taking into account the
$\alpha'$-corrections, one finds quantitatively different predictions
for the number density of D-dimensional black holes for both the
quantum and thermal interpretations. This can be seen easily seen if
one make use of the following relationship between the Euclidean
action and the entropy,
\begin{eqnarray}
S\;=\;(D-3) S_E\,+\,{\it O}(\alpha')\;.
\end{eqnarray}
Recall that $\rho=e^{S_E}$ while $\Omega=e^{S}$.

Our overall emerging picture of a quantum black hole is thus that of
an elementary excitation mode of some quantum p-brane living in its
geodesically complete regular spacetime condensate.

\acknowledgments

We are grateful to J.D.Bekenstein as well as the referees of a previous
version of the present manuscript for pointing out erroneous
results in our perturbative treatment as well as making us aware of
the work of Ref.[6].

This research was supported in part by the U.S. Department
of Energy under Grant No. DE-FG05-84ER40141.

\figure{Statistical Bootstrap : $\Omega(M)\,\sim\,\rho(M)\;\;;
(M \rightarrow \infty)\,$.\label{boot}}
\end{document}